\documentclass[twocolumn,aps,prl,superscriptaddress]{revtex4}

\usepackage{amsmath}
\usepackage{graphicx}
\usepackage{MnSymbol}

\usepackage[normalem]{ulem}
\usepackage{color}

\begin{document}

\title{Contact electrification and the work of adhesion}

\author{B.N.J. Persson}
\affiliation{Peter Gr\"unberg Institut-1, FZ-J\"ulich, 52425 J\"ulich, Germany}
\author{M. Scaraggi}
\affiliation{Peter Gr\"unberg Institut-1, FZ-J\"ulich, 52425 J\"ulich, Germany}
\affiliation{DII, Universita del Salento, 73100 Monteroni-Lecce, Italy}
\author{A.I. Volokitin}
\affiliation{Peter Gr\"unberg Institut-1, FZ-J\"ulich, 52425 J\"ulich, Germany}
\affiliation{Samara State Technical University, 443100 Samara, Russia}

\begin{abstract}
We present a general theory for the contribution from contact electrification to the work necessary to separate two solid bodies.
The theory depend on the surface charge density correlation function $\langle \sigma ({\bf x}) \sigma ({\bf 0})\rangle$
which we deduce from Kelvin Force Microscopy (KFM) maps of the surface electrostatic potential. For silicon
rubber (polydimethylsiloxane, PDMS) we discuss in detail the relative importance of the different contributions to the observed work of adhesion.
\end{abstract}

\maketitle

\pagestyle{empty}


When two solid objects are removed after adhesional or frictional contact, they will in general remain
charged\cite{Wan,Camara1,Camara2}. At the macroscopic level charging usually manifests itself as spark discharging upon contact with a third
(conducting) body, or as an adhesive force. The long-range electrostatic force resulting 
from charging is important in many technological
processes such as photocopying, laser printing, electrostatic separation methods, and sliding-triboelectric nanogenerators
based on in-plane charge separation\cite{Nano}. Contact charging is also the origin of
unwanted effects such as electric shocks, explosions or damage of electronic equipments.  

Contact electriﬁcation is one of the oldest areas of scientific
study, originating more than 2500 years ago when Thales of
Miletus carried out experiments showing that rubbing amber
against wool leads to electrostatic charging\cite{history}.
In spite of its historical nature and practical importance, there are many not well understood problems related to contact electrification, 
such as the role of surface roughness\cite{Persson,Robbins,PerssonA}, surface migration\cite{PerssonB} and contact de-electrification\cite{Soh}.

The influence of contact electrification on adhesion has been studied in pioneering work by Derjaguin et al\cite{Der1,Der2} and by Roberts\cite{Roberts}.
These studies, and most later studies, have assumed that removing the contact between two bodies results in the bodies having uniform surface charge
distributions of opposite sign. However, a very recent work\cite{Apod,Bayt,Baytekin} has shown that the bodies in general have surface charge distributions
which vary rapidly in space (on the sub-micrometer scale) between positive and negative values, and that the
net charge on each object is much smaller (sometimes by a factor of $\sim 1000$) than would result by integrating the absolute value of the 
fluctuating charge distribution over the surface area of a body.

Contact electrification occurs even between solids made from the same material\cite{Apod}. This has been demonstrated for
silicon rubber (PDMS). If two rubber sheets in adhesive contact (contact area $A$) are separated, they obtain net charges $\pm Q$ of opposite sign.
However, as discussed above, each surface has surface charge distributions fluctuating rapidly between positive and negative values,
with magnitudes much higher than the average surface charge densities $\pm Q/A$. The net charge
scales with the contact surface area as $Q\sim A^{1/2}$, as expected based on a picture where the net charge results from randomly
adding positively and negatively charged domains (with individual area $\Delta A$) on the surface area $A$: when $N=A/\Delta A >> 1$,
we expect from statistical mechanics that the net charge on the surface $A$ is proportional to $N^{1/2}$ as observed\cite{Apod}. Note that in
the thermodynamic limit, $A\rightarrow \infty$, the net surface charge density $Q/A = 0$.

In this letter we will present an accurate calculation of the contribution from contact electrification to the work of adhesion to separate
two solids. The same problem has been addressed in a less accurate approach by Br\"ormann et al\cite{Bror}. They assumed that the charged
domains formed a mosaic pattern of squares, where each nearby square has charge of opposite sign but of equal magnitude. To this problem
they applied an approximate procedure to obtain the contribution to the work of adhesion from charging. In this letter
we will present a general theory, where the surface charge distribution $\sigma ({\bf x})$ is characterized by the density-density correlation function 
$\langle \sigma ({\bf x})\sigma ({\bf 0})\rangle$, the power spectrum of which can be deduced directly from
Kelvin Force Microscopy (KFM) potential maps. 
We find that for polymers the contact electrification may contribute a non-negligible amount to the observed 
work of adhesion. However, more KFM measurements at smaller tip-substrate separation are necessary to confirm the conclusion presented below.

\begin{figure}
\includegraphics[width=0.7\columnwidth]{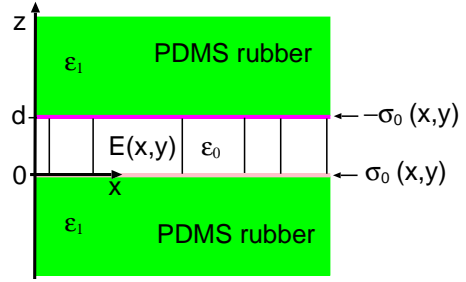}
\caption{\label{pic1}
After separation the bottom solid has the surface charge distribution $\sigma_0 ({\bf x})$ and the top
solid the surface charge distribution $-\sigma_0 ({\bf x})$. 
}
\end{figure}

We will calculate the force between the two charged solids when the surfaces are separated by the distance $d$, see Fig. \ref{pic1}.
The lower surface has the surface charge density $\sigma_0({\bf x})$, where ${\bf x} = (x,y)$ is the in-plane coordinate, and the upper surface the surface charge density $-\sigma_0 ({\bf x})$.
We write the electric field as ${\bf E} = -\nabla \phi$ so that the electric potential $\phi$ 
satisfies $\nabla^2 \phi = 0$ everywhere except for $z=0$ and $z=d$.
We write
$$\sigma_0 ({\bf x}) = \int d^2q \ \sigma_0 ({\bf q}) e^{i{\bf q}\cdot {\bf x}}.$$
The electrostatic stress tensor
$$\sigma_{ij} = {1\over 4 \pi} \left (E_i E_j -{1\over 2} {\bf E}^2 \delta_{ij}\right ).$$
Here we are interested in the $zz$-component:
$$\sigma_{zz} = {1\over 8 \pi} \left (E_z^2-{\bf E}_\parallel^2\right ).\eqno(1)$$
In the space between the surfaces the electric potential:
$$\phi = \int d^2q \left [\phi_0({\bf q}) e^{-qz}+ \phi_1({\bf q}) e^{qz} \right ] e^{i{\bf q}\cdot {\bf x}}$$
where ${\bf q} = (q_x,q_y)$ and ${\bf x} = (x,y)$ are 2D vectors. Thus for $z=0$:
$$E_z = \int d^2 q  \ q \left [ \phi_0({\bf q}) - \phi_1({\bf q}) \right ] e^{i{\bf q}\cdot {\bf x}}\eqno(2)$$
and
$${\bf E}_\parallel = \int d^2 q  (-i{\bf q}) \left [\phi_0({\bf q}) + \phi_1({\bf q}) \right ] e^{i{\bf q}\cdot {\bf x}}.\eqno(3)$$
Using (1), (2) and (3) gives
$$\int d^2 x \ \sigma_{zz} = 2 \pi {\rm Re} \int d^2 q \ q^2 \phi_0({\bf q}) \phi^*_1({\bf q}).\eqno(4)$$
We now calculate $\phi_0({\bf q})$ and $\phi_1({\bf q})$. We write the electric
potential $\phi({\bf q},z)$ as:
$$\phi = \phi_0 e^{-qz}+ \phi_1 e^{qz} \ \ \ \ \ {\rm for}  \ \ 0<z<d,$$
$$\phi = \phi_2 e^{qz} \ \ \ \ \ {\rm for}  \ \ z<0,$$
$$\phi = \phi_3 e^{-q(z-d)} \ \ \ \ \ {\rm for}  \ \ z>d.$$
Since $\phi$ must be continuous for $z=0$ and $z=d$ we get:
$$\phi_0+ \phi_1 = \phi_2\eqno(5)$$
$$\phi_0 e^{-qd}+\phi_1 e^{qd} = \phi_3.\eqno(6)$$
Let $\epsilon_0$ and $\epsilon_1$ be the dielectric function of the region between the bodies ($0<z<d$) and in the bodies ($z<0$ and $z>d$),
respectively. In our application the space between the bodies is filled with non-polar gas and $\epsilon_0 \approx 1$.
From the boundary conditions $\epsilon_0 E_z(0^+)-\epsilon_1 E_z(-0^+) = 4 \pi \sigma_0$ and  
$\epsilon_1 E_z(d+0^+)- \epsilon_0 E_z(d-0^+) = - 4 \pi \sigma_0$, 
and using (5) and (6), we get:
$$\phi_0 + g \phi_1 = {2\pi \over q} \sigma$$
$$g \phi_0 e^{-qd}+\phi_1 e^{qd} = - {2\pi \over q} \sigma$$
where $\sigma = \sigma_0 2 / (\epsilon_1+\epsilon_0)$ and $g= (\epsilon_1-\epsilon_0)/(\epsilon_1+\epsilon_0)$.
Solving these equations gives:
$$\phi_0 = {2\pi \over q}  { \sigma \over 1+ge^{-qd}}, \ \ \ \ \ \ \phi_1 = e^{-qd} \phi_0.$$
Using these equations in (4) gives
$$\langle F_z \rangle =\int d^2 x \ \langle \sigma_{zz} \rangle = (2 \pi )^3 \int d^2 q  \ \langle | \sigma({\bf q})|^2 \rangle {e^{-qd} 
\over \left (1+ge^{-qd} \right )^2} \eqno(7)$$
where we have performed an ensemble average denoted by $\langle .. \rangle$.

Consider the correlation function:
$$\langle | \sigma ({\bf q})|^2 \rangle ={1\over (2\pi)^4} \int d^2x d^2x' \langle  \sigma ({\bf x})  \sigma ({\bf x}') \rangle e^{i{\bf q}\cdot ({\bf x}-{\bf x}')}.$$
Assuming that the statistical properties of the surface charge distribution are translational invariant we get:
$$ \langle  \sigma ({\bf x})  \sigma ({\bf x}') \rangle =  \langle  \sigma ({\bf x}-{\bf x}')  \sigma ({\bf 0}) \rangle$$
and 
$$\langle | \sigma ({\bf q})|^2 \rangle ={A_0\over (2\pi)^4} \int d^2x  \langle \sigma ({\bf x}) \sigma ({\bf 0}) \rangle e^{i{\bf q}\cdot {\bf x}}$$
where $A_0$ is the surface area. If $\bar \sigma = \langle  \sigma ({\bf x}) \rangle$ denote the average surface charge density, then we define the charge density
power spectrum:
$$C_{\sigma \sigma} ({\bf q}) = {1\over (2\pi )^2} \int d^2 x  \langle [ \sigma ({\bf x})-\bar \sigma] [ \sigma ({\bf 0})-\bar \sigma] \rangle e^{i{\bf q}\cdot {\bf x}}.\eqno(8)$$ 
Using this definition we get:
$$\langle | \sigma ({\bf q})|^2 \rangle ={A_0\over (2\pi)^2} \left [ C_{\sigma \sigma} ({\bf q}) + \bar \sigma^2 \delta ({\bf q}) \right ].\eqno(9)$$
Substituting (9) in (7) gives
$$\langle F_z \rangle = 2 \pi A_0 \bar \sigma^2 +  2 \pi A_0 \int d^2 q \ C_{\sigma \sigma}({\bf q})  {e^{-qd} \over \left (1+ge^{-qd} \right )^2}.$$
We expect the statistical properties of the surface charge distribution to be isotropic which imply that $C_{\sigma \sigma}({\bf q})$ only depends on the magnitude
$q=|{\bf q}|$. This gives:
$$\langle F_z (d) \rangle = 2 \pi A_0 \bar \sigma^2 +  (2 \pi)^2 A_0 \int d q \ q C_{\sigma \sigma}(q) {e^{-qd} \over \left (1+ge^{-qd} \right )^2}.$$
The first term in this expression is the attraction between the surfaces due to the (average) uniform component of the charge distribution which, as expected, is independent
of the separation between the surfaces (similar to a parallel condenser). The second term is the contribution from the fluctuating components of the surface charge
distribution. 
The contribution to the work of adhesion from the surface charge is given by:
$$U = \int_0^d dz \ \langle F_z(z) \rangle 
=  2 \pi A_0 \bar \sigma^2 d$$
$$ +  (2 \pi)^2 A_0 \int_0^\infty d q \ q C_{\sigma \sigma}(q) \int_0^d dz {e^{-qz} \over \left (1+ge^{-qz} \right )^2}.\eqno(10)$$ 
The first term increases without limit as the surfaces are separated, and we will not include this term in the work of adhesion. For bodies of finite size the expression given above
for the contribution from the net charging is of course only valid for separations smaller than the linear size of the bodies (i.e. $d < L$, where $A_0=L^2$). The contribution
to the work of adhesion from the second term in (10) (for $d \rightarrow \infty$) is:
$$w_{\rm ch} = {U \over A_0} = {(2 \pi)^2 \over 1+g}  \int_0^\infty d q \ C_{\sigma \sigma}(q). \eqno(11)$$

Note that the integral
$$\int d^2q \ C_{\sigma \sigma}({\bf q}) = \langle [ \sigma ({\bf x})-\bar \sigma]^2 \rangle = \langle \Delta \sigma^2 \rangle \eqno(12)$$
is the mean of the square of the fluctuating surface charge distribution. 
Using this equation we can write:
$$w_{\rm ch} = {2 \pi \over 1+g} {\langle \Delta \sigma^2\rangle \over \langle q \rangle} \eqno(13)$$
where
$$\langle q \rangle  =   {\int_0^\infty d q \ q C_{\sigma \sigma}(q) \over \int_0^\infty d q \  C_{\sigma \sigma}(q)}.\eqno(14) $$

The study above is for the limiting case where the surfaces separate so fast that no decay in the surface charge distribution
takes place before the separation is so large as to give a negligible interaction force. Experiments\cite{Bayt} have shown that the
charge distribution decay with increasing time as ${\rm exp} (-t/\tau)$, where the relaxation time $\tau \approx 10^3 \ {\rm s}$
depends on the atmospheric condition (e.g., humidity and concentration of ions in the surrounding gas). Taking into account the decay in the surface
charge distribution, and assuming $z=vt$ (where $v$ is the normal separation velocity) we need to replace the integral over $z$ in (10) with:
$$ f(q,v) = \int_0^\infty  dz {e^{-qz} e^{-2t/\tau} \over \left (1+ge^{-qz} \right )^2} = 
\int_0^\infty  dz {e^{-(qz +2z/v\tau )}  \over \left (1+ge^{-qz} \right )^2}$$ 
and (11) becomes
$$w_{\rm ch} = (2 \pi)^2   \int_0^\infty d q \ q C_{\sigma \sigma}(q) f(q,v). \eqno(15)$$
In the limit $v \rightarrow \infty$ we have $f\rightarrow 1/[q(1+g)]$ and in this limit (15) reduces to (13).
In the opposite limit of very small surface separation velocity, $f \rightarrow v \tau /[2(1+g)^2]$ and in this limit
$$w_{\rm ch} = {(2 \pi)^2 v \tau \over 2(1+g)^2}   \int_0^\infty d q \ q C_{\sigma \sigma}(q) = {\pi  v \tau \langle \Delta \sigma^2\rangle \over (1+g)^2}.\eqno(16)$$
Note that this expression is of the form (13) with $1/\langle q \rangle$ replaced by $v \tau/[2(1+g)]$. Since typically $\tau \approx 10^3 \ {\rm s}$
and $(1+g) \approx 1$ and $\langle q \rangle \approx q_1 \approx 10^9 \ {\rm m}^{-1}$ (where $q_1$ is defined below) 
we get $v_{\rm c} = 2 (1+g)/(\langle q \rangle \tau) \approx 10^{-12} \ {\rm m/s}$.
In most applications we expect the separation velocity in the vicinity of the crack tip $v >> v_{\rm c}$, and in this case the limiting equation (13)
holds accurately. Note, however, that the separation velocity $v$ may be much smaller than the crack tip velocity. 

In the KFM measurement the local potential at some fixed distance $d$ above the surface is measured, rather than the surface charge density. From the measured data the 
potential power spectrum 
$$C_{\phi \phi} ({\bf q}) = {1\over (2\pi )^2} \int d^2 x  \langle [\phi ({\bf x})-\bar \phi] [\phi ({\bf 0})-\bar \phi] \rangle e^{i{\bf q}\cdot {\bf x}}$$
can be directly obtained. However, we can relate the potential to the charge density:
$$\phi ({\bf q}) = {2 \pi \over q} \sigma ({\bf q}) e^{-qd}.$$
Thus
$$C_{\sigma \sigma}({\bf q}) = {q^2 \over (2 \pi )^2} C_{\phi \phi} ({\bf q}) e^{2qd}.\eqno(17)$$

The results presented above is in Gaussian units. To obtain (17) in SI units we must multiply
the right-hand-side with $(4\pi \epsilon_0)^2$, where $\epsilon_0 = 8.8542 \times 10^{-12} \ {\rm C V^{-1} m^{-1}}$.
Thus: 
$$C_{\sigma \sigma}({\bf q}) = 4 \epsilon_0^2 q^2  C_{\phi \phi} ({\bf q}) e^{2qd}.\eqno(18)$$
To get (11) in SI units we must multiply the right-hand-side by $(4 \pi \epsilon_0)^{-1}$:
$$w_{\rm ch} = {\pi \over 2\epsilon_0 (1+g)}  \int_0^\infty d q \ C_{\sigma \sigma}(q). \eqno(19)$$

\begin{figure}
\includegraphics[width=0.8\columnwidth]{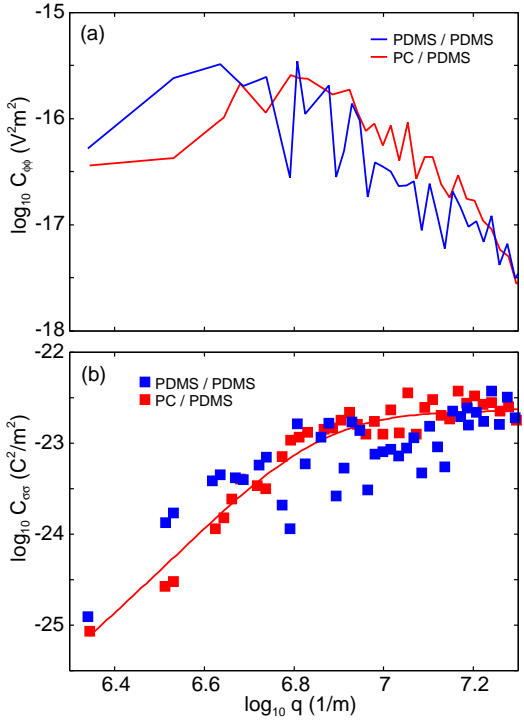}
\caption{\label{both}
(a) The voltage power spectrum $C_{\phi \phi}$ and (b) the surface charge density power spectrum $C_{\sigma \sigma}$
as a function of the wavevector. The results have been calculated from the measured (KFM)
voltage maps for PDMS/PDMS (blue) and PDMS/polycarbonate (PC) (red)\cite{Bayt}.
}
\end{figure}

We now analyze experimental data involving elastically soft solids with smooth surfaces, where the initial contact between the solids is 
complete due to the adhesion between the solids. In Ref. \cite{Bayt} several such systems where studied
and here we focus on PDMS rubber against PDMS. After breaking the adhesive contact between two sheets of PDMS 
(which involves interfacial crack propagation) the electrostatic potential a distance $d$ above one of the surfaces was
probed using KFM measurements. From the measured potential map we have calculated the potential power spectrum
$C_{\phi \phi}(q)$ and then from (18) the charge density power spectrum $C_{\sigma \sigma}(q)$. The measurements where done at the tip-substrate
separation $d\approx 10^{-7} \ {\rm m}$, and since the electric potential from a surface charge density distribution with the
wavevector $q$ decay as ${\rm exp}(-qd)$ with the distance $d$ from the surface, the KFM is effectively limited to probing the surface
charge distribution with wavevector $q < 1/d$. In Fig. \ref{both} we show both power spectra's for $q < 2\times 10^7 \ {\rm m^{-1}}$.
Note that the charge density power spectrum appears to saturate for large wavevector, say $q>q_0$, with $q_0 \approx 10^7 \ {\rm m}^{-1}$.
This result follows if, as expected, the process of creating surface charges is uncorrelated in space at short length scales. In that case
$\langle \sigma({\bf x}) \sigma ({\bf 0}) \rangle \sim \delta ({\bf x})$ and using (8) this gives $C_{\sigma \sigma} ({\bf q}) = {\rm const}$. 
The fact that $C_{\sigma \sigma} ({\bf q})$ decays for decreasing $q$ for $q < q_0 \approx 10^7 \ {\rm m}^{-1}$ implies that at some length scale
$\lambda_0 = 2 \pi /q_0 \approx 0.6 \ {\rm \mu m}$ the charge distribution becomes correlated. The physical reason for this may relate to inhomogenities
on the PDMS surface, e.g., domains of slightly varying PDMS composition or cross-linking density. (Note: PDMS rubber is obtained by mixing two high viscosity liquids
and may exhibit inhomogeneties at the micrometer scale, e.g., due to incomplete mixing.)

We assume that the charge density power spectrum saturate for $q>q_0$ at $C^0_{\sigma \sigma} \approx 2.2\times 10^{-23} \ {\rm C^2/m^2}$ (see Fig. \ref{both}(b)).
In this case from (19) we get $w_{\rm ch} \approx (q_1-q_0)C^0_{\sigma \sigma} /\epsilon_0$, where $q_1$ is a large wavevector cut-off of order $2\pi /\lambda_1$, where
$\lambda_1$ is of order the average separation between the surface charges (which we assume to be point charges of magnitude $\pm e$, where $e$ is the electron charge). 
Here we have used that $\pi /[2(1+g)] \approx 1$. Since $q_0 \approx 10^7 \ {\rm m^{-1}} << q_1$ we get 
$w_{\rm ch} \approx q_1 C^0_{\sigma \sigma} /\epsilon_0 \approx 0.002 \ {\rm J/m^2}$, where we have assumed $q_1 = 10^9 \ {\rm m}^{-1}$.
This value is smaller than the measured work of adhesion during adiabatic (very slow) separation of the surfaces where $w_{\rm ch} \approx 0.05  \ {\rm J/m^2}$.
Using (12) we get the mean square charge fluctuation $\langle \Delta \sigma^2\rangle  \approx \pi q_1^2 C^0_{\sigma \sigma} \approx 7\times 10^{-5} \ {\rm C^2/m^4}$ or
the rms charge fluctuation $\approx 1 \ {\rm \mu C/cm^2}$ which is similar to what was estimated by Baytekin et al\cite{Bayt}.

The analysis above is based on the assumption that the surface charge density power spectrum saturates at a value 
$C^0_{\sigma \sigma} \approx 2.2\times 10^{-23} \ {\rm C^2/m^2}$ for large wavevectors, and that the cut-off $q_1 \approx 10^9 \ {\rm m}^{-1}$, corresponding to an
average separation between the point charges of about $6 \ {\rm nm}$. This hypothesis should be tested by performing
KFM measurements to smaller tip-substrate separations. The number of surface charges, which determines the
cut-off $q_1$ in the study above, may also be probed by surface reaction experiments, such as bleaching experiments reported on in Ref. \cite{Baytekin}. 

The value of $q_1$ used above corresponds to $\lambda_1^{-2} \approx 3 \times 10^{16}$ electrons per ${\rm m}^{-2}$. If these charges
would result from breaking of the PDMS polymer chains, it would require at least $(3 \times 10^{16} \ {\rm m}^{-2}) \times (3 \ {\rm eV} ) \approx 0.015 \ {\rm J/m^2}$,
which is smaller than the observed work of adhesion, but not negligible. For PDMS the observed work of adhesion at low separation 
velocity  equals $w \approx 0.05 \ {\rm J/m^2}$). At low crack-tip velocities the viscoelastic energy dissipation at the crack tip, 
and other non-equilibrium effects are negligible, the work of adhesion is usually assumed to result from the van der Waals interaction between
the surfaces at the interface, but the study above indicate that there may be 
non-negligible contributions both from the bond-breaking process which generates the surface charges, and from
the contact electrification itself. 

To summarize, we have derived a general expression for the contribution to the work of adhesion from contact electrification, and 
we have shown that for PDMS (and for polymers in general) the contact electrification and the associated bond-breaking may contribute in a non-negligible way to the observed 
work of adhesion.

\vskip 0.3cm
{\bf Acknowledgments}
We thank B. Baytekin, H.T. Baytekin and  B.A. Grzybowski for kindly supplying the KFM potential maps used in calculating
the power power spectras shown in Fig. 2. A.I.V. acknowledges financial support from
Russian Foundation for Basic Research (Grant N 12-02-00061-a). M. Scaraggi acknowledge support from FZ J\"ulich.

\end{document}